\documentclass[lettersize,journal,twoside]{IEEEtran}

\usepackage{amsmath,amssymb,amsfonts, amsthm}
\usepackage{bm}
\usepackage{epsfig}
\usepackage[utf8]{inputenc} 
\usepackage{amsbsy} % gives \boldsymbol,\pmb
\usepackage[dvipsnames]{xcolor}
\usepackage[table]{xcolor}
\usepackage{acronym}
\usepackage{soul}

\usepackage{cite}
\usepackage{color}
\usepackage{graphicx}
\usepackage{tcolorbox}
\def\blfootnote{\xdef\@thefnmark{}\@footnotetext}
\makeatother

\makeatletter
\def\blfootnote{\xdef\@thefnmark{}\@footnotetext}
\makeatother

\newcommand{\textjj}[1]{\textcolor{black}{#1}}

\acrodef{3GPP}[3GPP]{3rd Generation Partnership Project}
\acrodef{5G}[5G]{fifth-generation}
\acrodef{6G}[6G]{sixth-generation}
\acrodef{ADC}[ADC]{Analog-to-Digital Converter}
\acrodef{AO}[AO]{Alternating Optimization}
\acrodef{AoA}[AoA]{Angle of Arrival}
\acrodef{AoD}[AoD]{Angle of Departure}
\acrodef{APN}[APN]{Analog Precoding Network}
\acrodef{ARV}[ARV]{Array Response Vector}
\acrodef{ASK}[ASK]{Amplitude-Shift Keying}
\acrodef{AWGN}[AWGN]{Additive White Gaussian Noise}
\acrodef{BER}[BER]{Bit Error Ratio}
\acrodef{BF}[BF]{Beamforming}
\acrodef{BFN}[BFN]{Beamforming Network}
\acrodef{BS}[BS]{Base Station}
\acrodef{CB}[CB]{conjugate beamforming}
\acrodef{COMP}[COMP]{Covariance OMP}
\newacro{CSI}{channel state information}
\newacro{CUMA}{Compact Ultra Massive Antenna Array}
\acrodef{DAC}[DAC]{Digital to Analog Converter}
\acrodef{DAS}[DAS]{distributed antenna system}
\acrodef{DBS}[DBS]{Distance-Based Scheduling}
\acrodef{DCS}[DCS]{Digital Communication System}
\acrodef{DCOMP}[DCOMP]{Dynamic COMP}
\newacro{DFT}{Discrete Fourier Transform}
\acrodef{DC}[DC]{Digital Combining}
\acrodef{DL}[DL]{downlink}
\acrodef{DOA}[DOA]{Direction Of Arrival}
\acrodef{DoF}[DoF]{degrees-of-freedom}
\acrodef{DPC}[DPC]{dirty-paper coding}
\acrodef{ELAA}[ELAA]{extremely-large aperture array}
\acrodef{ESD}[ESD]{Energy Spectral Density}
\acrodef{FAMA}[FAMA]{Fluid Antenna Multiple Access}
\acrodef{FCC}[FCC]{Federal Communications Commission}
\newacro{FDD}{Frequency-Division Duplex}
\acrodef{FF}{far-field}
\acrodef{FR}{Frequency Range}
\acrodef{FLS}[FLS]{Front Line Scheduling}
\acrodef{FSK}[FSK]{Frequency-Shift Keying}
\acrodef{FT}[FT]{Fourier Transform}
\acrodef{GEPort}[GEPort]{Generalized Eigenvector Port Selection}
\acrodef{HT}[HT]{Hilbert Transform}
\acrodef{HI}[HI]{Harmonic Interference}
\acrodef{ICI}[ICI]{Inter-Carrier Interference}
\acrodef{gMIMO}[gMIMO]{gigantic MIMO}
\acrodef{IL}[IL]{Insertion Losses}
\acrodef{ISI}[ISI]{Inter-Symbol Interference}
\acrodef{ITU}[ITU]{International Telecommunications Union}
\acrodef{IUI}[IUI]{inter-user interference}
\acrodef{JSDM}[JSDM]{Joint Spatial Division and Multiplexing}
\acrodef{LBFN}[LBFN]{Linear Beamforming Network}
\acrodef{LLF}[LLF]{Log-Likelihood function}
\acrodef{LMD}[LMD]{Linearly Modulated Digital}
\acrodef{LoS}[LoS]{Line-of-Sight}
\newacro{MA}{Modular Array}
\newacro{MAP}{Maximum A Posteriori}
\acrodef{MIMO}[MIMO]{Multiple-Input Multiple-Output}
\newacro{ML}{Maximum Likelihood}
\newacro{MMSE}{Minimum Mean Squared Error}
\acrodef{MMV}[MMV]{Multiple Measurement Vector}
\acrodef{mmWave}[mmWave]{millimeter wave}
\acrodef{MRC}[MRC]{Maximum Ratio Combining}
\acrodef{MRT}[MRT]{Maximum Ratio Transmission}
\newacro{MSE}{Mean Squared Error}
\acrodef{MUSIC}[MUSIC]{MUltiple SIgnal Classification}
\acrodef{NF}{near-field}
\acrodef{NLoS}[NLoS]{Non Line-of-sight}
\acrodef{NMSE}{Normalized Mean Squared Error}
\acrodef{OFDM}[OFDM]{Orthogonal Frequency-Division Multiplexing}
\acrodef{OFDMA}[OFDMA]{Orthogonal Frequency-Division Multiple Access}
\acrodef{OMP}[OMP]{Orthogonal Matching Pursuit}
\acrodef{OSMP}[OSMP]{Orthogonal Subspace Matching Pursuit}
\acrodef{PA}[PA]{Power Amplifier}
\acrodef{PS}[PS]{Phase Shifter}
\acrodef{PSK}[PSK]{Phase-Shift Keying}
\acrodef{PW}{planar wavefront}
\acrodef{QAM}[QAM]{Quadrature Amplitude Modulation}
\acrodef{RF}[RF]{Radio Frequency}
\acrodef{RFC}[RFC]{Rayleigh Fading Channel}
\acrodef{RSS}[RSS]{Rectangular-Search Scheduler}
\acrodef{SDMA}[SDMA]{space-division multiple access}
\acrodef{SE}[SE]{spectral efficiency}
\acrodef{SER}[SER]{Symbol Error Rate}
\acrodef{SIC}[SIC]{Successive Interference Cancellation}
\acrodef{SR}[SR]{Sideband Radiation}
\acrodef{SINR}[SINR]{signal-to-interference-plus-noise ratio}
\acrodef{SLL}[SLL]{Side-Lobe Level}
\acrodef{SOCP}[SOCP]{Second-Order Cone Program}
\acrodef{SOMP}[SOMP]{Simultaneous-Orthogonal Matching Pursuit}
\acrodef{SPDT}[SPDT]{Single-Pole-Double-Throw}
\acrodef{SPST}[SPST]{Single-Pole-Single-Throw}
\acrodef{SR}[SR]{Sideband Radiation}
\acrodef{SS}[SS]{Spatial Smoothing}
\acrodef{SNR}[SNR]{Signal-to-Noise Ratio}
\acrodef{SUS}{successive user selection}
\acrodef{SW}{spherical wavefront}
\newacro{TDD}{Time-Division Duplex}
\acrodef{TM}[TM]{Time Modulation}
\acrodef{TMA}[TMA]{Time-Modulated Array}
\acrodef{ULA}[ULA]{Uniform Linear Array}
\acrodef{UM-MIMO}[UM-MIMO]{Ultra Massive MIMO}
\acrodef{UPA}[UPA]{Uniform Planar Array}
\acrodef{UPW}[UPW]{Uniform Planar Wave}
\acrodef{USW}[USW]{Uniform Spherical Wave}
\acrodef{VGA}[VGA]{Variable Gain Amplifier}
\acrodef{VPS}[VPS]{Variable Phase Shifter}
\acrodef{VR}[VR]{visibility regions}
\acrodef{XL}[XL]{extra-large}
\acrodef{XL-array}[XL-array]{extra-large array}
\acrodef{XL-MIMO}[XL-MIMO]{extra-large Multiple-Input Multiple-Output}
\acrodef{XL-ULA}[XL-ULA]{extra-large uniform linear array}
\acrodef{ZF}[ZF]{Zero-Forcing}

\begin{document}

\title{\LARGE{Channel Estimation and Reconstruction in Fluid Antenna Multiple Access: Myths, Misconceptions and Critical Questions}}

\author{Taissir Y. Elganimi, Pablo Ram\'irez-Espinosa, David Morales-Jim\'enez and F. Javier L\'opez-Mart\'inez
\thanks{Manuscript received May 31, 2026; revised Month XX, 2026; accepted Month XX, 2026. This work has been submitted to the IEEE for publication. Copyright may be transferred without notice, after which this version may no longer be accessible.}
\thanks{T. Y. Elganimi, F.J. L\'opez-Mart\'inez and D. Morales-Jim\'enez are with the Department of Signal Theory, Networking and Communications, Research Centre for Information and Communication Technologies (CITIC-UGR), University of Granada, 18071, Granada, Spain. P. Ram\'irez-Espinosa is with Telecommunications Research Institute (TELMA), University of M\'alaga, M\'alaga 29071 (Spain).} %(contact: \texttt{etaissir@ugr.es}).}

%\thanks{The work of P.R.E is funded by the European Union under the Marie Sklodowska-Curie grant agreement No. 101109529. The work of T.Y.E, F.J.L.M. and D.M.J. is supported in part by the State Research Agency (AEI) of Spain and the European Social Fund under grant RYC2020-030536-I, by MICIU/AEI/10.13039/501100011033 and FEDER/UE under grant PID2023-149975OB-I00 (COSTUME), and by PREP2023-001949 grant.}
}
\markboth{submitted to IEEE}{Channel Estimation and Reconstruction in Fluid Antenna Multiple Access\ldots}

\maketitle

\begin{abstract}
\textjj{Fluid antenna systems (FAS) represent a paradigm shift in which antenna elements (ports) emulate the illusion of motion or fluidity within a spatial aperture to optimize performance. One of FAS's key use cases is the provision of open-loop fluid antenna multiple access (FAMA), enabling multiplexing gains through spatial interference nulling without requiring channel state information (CSI) at the transmitter side. However, this comes at the price of requiring a precise channel reconstruction at the receiver to successfully identify the optimal port. Current research efforts map this sensing task to a legacy MIMO-style estimation problem focused on minimizing global reconstruction errors such as normalized mean-squared error (NMSE). In this work, we argue that because FAS is inherently selection-based, NMSE-like approaches often lead to excessive training overhead and reduced net throughput. We revisit the problem of channel estimation and reconstruction in FAS, challenging some prevalent myths related to (\textit{i}) the adequacy of global error metrics; (\textit{ii}) the convenience of reconstructing channels or aggregate interference; (\textit{iii}) the need for spatial oversampling; and (\textit{iv}) the impact of port selection accuracy. We also identify four critical questions that must be answered for successfully enabling FAMA deployments: (\textit{i}) the definition of a selection-optimal sampling law; (\textit{ii}) the identification of proper reconstruction methodologies; (\textit{iii}) the inherent trade-offs between multi-port sensing and selection gain; and (\textit{iv}) the challenges introduced when moving towards electronically reconfigurable FAS.}

\end{abstract}


%---------------------------------------------
% Section I - INTRODUCTION 
%---------------------------------------------
\section{Introduction}

Fluid antenna systems (FAS) refer to any architecture where antenna elements can be moved (either physically or virtually) within a fixed spatial aperture \cite{Tutorial2025}. FAS have emerged as a compellingly simple yet convenient alternative to traditional massive multiple-input multiple-output (MIMO) systems, offering promising multiplexing capabilities while relaxing many of the complexity burdens associated with the latter. By leveraging dynamic reconfigurability to minimize interference and exploit spatial diversity, early FAS-based multiple access solutions such as slow fluid antenna multiple access (FAMA) \cite{Wong2023} enable multi-user connectivity without the need for sophisticated signal processing at the transmitter or receiver sides.

One of the key benefits of slow-FAMA lies in its open-loop operation, avoiding the need for \ac{CSI} at the transmitter side. However, this apparent advantage becomes highly asymmetric at the receiver: to fully unlock the spatial interference nulling gains that genuinely define FAMA, a precise spatial reconstruction of the \ac{SINR} is needed to determine the set of ports that yield the maximum \ac{SINR} \cite{Chai2022}. This requires accurate CSI across the fluid antenna (FA) aperture for both the desired and interfering channels.

Unlike conventional MIMO systems, FAS deploy their ports with sub-wavelength spatial resolution. Consequently, for a finite aperture size, the CSI acquisition problem can be mapped to a spatial sampling and reconstruction task where a number of samples (ideally much lower than the number of ports) is used to estimate the channels. This problem has been addressed in the literature from various perspectives \cite{New2025,Xu2025,Kang2026,Zhang2025,Liang2025}, often relying on channel estimation and reconstruction protocols that share a common basis with legacy MIMO systems in terms of pilot overhead and performance evaluation. However, as we will argue later, FAs are inherently selection-based systems. As such, they only require satisfactory reconstruction at the ports likely to be selected, ensuring that the performance loss due to sub-optimal port selection is minimized. In this work, we make the case that these unique features require a fundamental shift in perspective. To this end, we rebut and clarify a number of myths and misconceptions related to FAS channel estimation and reconstruction, and identify four critical questions that must be answered for FA-based multiple access techniques to become a successful contender in next generation wireless technologies.

\begin{figure*}[t]
\centering
{\includegraphics[width=2\columnwidth]{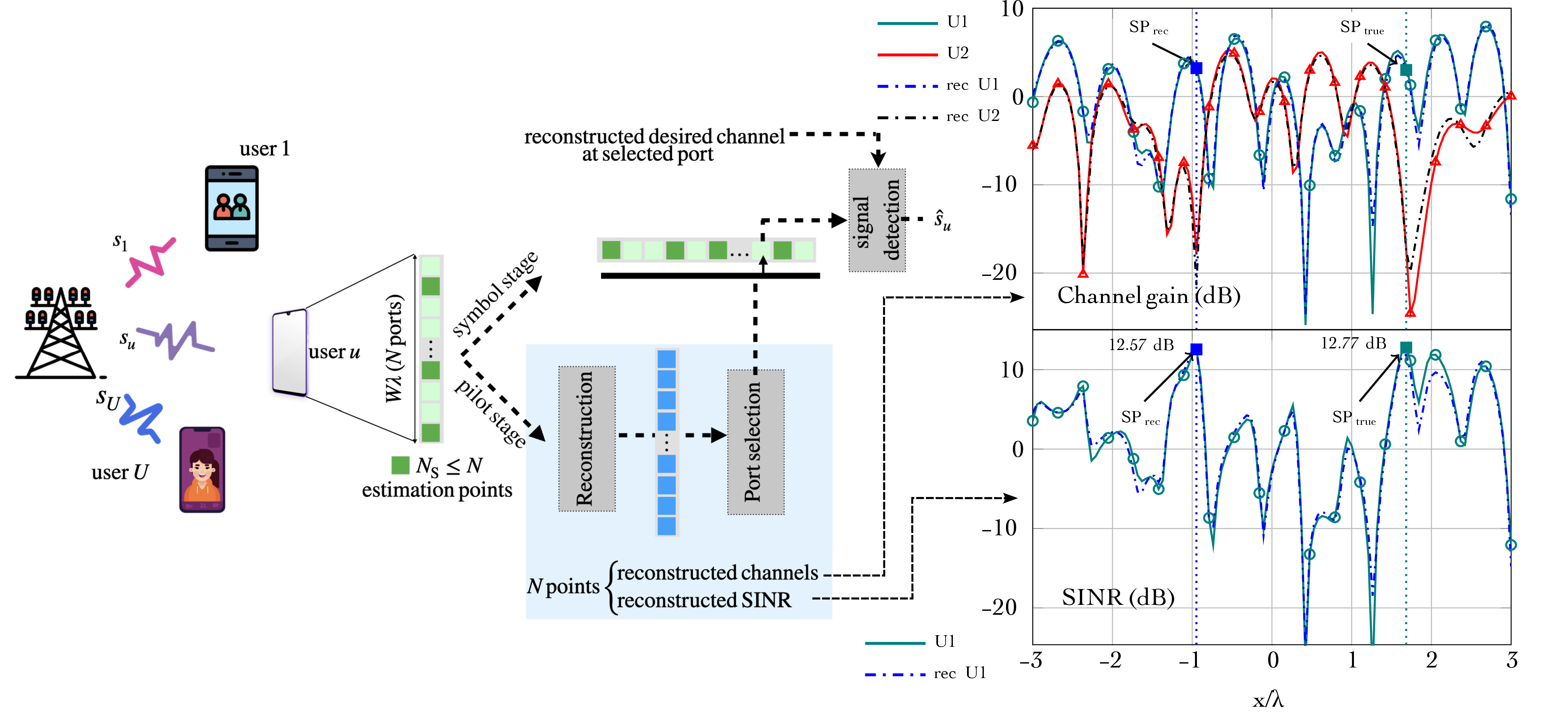}}
\caption{System model for the downlink of a $U$-user slow-FAMA set-up \cite{Wong2023}. For representation simplicity, a 1D FA is represented. At each receiver, the channels are estimated at some $N_{\rm S}$ points and then used to reconstruct the SINR values at all $N$ ports. In a toy example, the channel gains and resulting SINRs across the FA aperture at user 1 are represented, using a FA of $W=6$ wavelengths. Empty markers correspond to the $N_{\rm S}$ estimation points, with a sampling density of 3 points per wavelength.}
\label{figure_1}
\end{figure*}

%---------------------------------------------
% Section II - Channele estimation in FAS vs MIMO (legacy) 
%---------------------------------------------
\section{Channel estimation and reconstruction in FAS}

Let us consider the system model for FAMA depicted in Fig. \ref{figure_1}. According to the general definition in \cite{Wong2023}, the BS is equipped with $U$ antennas\footnote{This can be generalized to a system with $N_{\rm T}>>U$ antennas implementing random beamforming, to ensure equivalent conditions to that of a rich scattering environment. In this circumstances, the system model becomes equivalent to that in Fig. \ref{figure_1}.} that give service to $U$ users, each of which is equipped with a FA. The system operates in an open-loop fashion, so that CSI availability at the transmitter side is not required. The BS transmits the signal $s_u$ making use of the $u$-th transmit antenna, with $u=1, 2, \ldots U$. At the receiver side, each user receives a copy of all the signals generated at the BS, and aims to retrieve the desired signal $s_u$ that will be affected by interference coming from $s_{\tilde{u}}$ for $ \tilde{u} \neq u$. Since users are equipped with FAs with $N$ ports, each user can select the FA configuration that is more beneficial for their interest. Classically, this is accomplished by selecting the FA port $n^*\in[1, 2, \ldots N]$ that retrieves the largest \ac{SINR}; this is referred to as port selection \cite{Chai2022}.

The problem of port selection requires that user $u$ has \ac{CSI} knowledge for the set of desired channels (from the $u$-th transmit antenna to its $n$-th FA port), and also for the set of interfering channels  (from the $\tilde{u}$-th transmit antenna with $ \tilde{u} \neq u$ to its $n$-th FA port), such that the \ac{SINR} at all $N$ ports can be reconstructed to enable port selection, ideally from a much lower number of observations $N_{\rm S}$.

Existing channel estimation and reconstruction methods in FAS generally follow the same principles adopted in traditional MIMO systems. Specifically, channel estimation and reconstruction are typically performed by first transmitting orthogonal pilot sequences in the downlink, one per user, for the sake of estimating the channels of interest at the $N_{\rm S}$ sampling positions. This is followed by reconstructing the channels at all $N$ FA ports using spatial reconstruction techniques such as filtering interpolation \cite{New2025}, Bayesian methods \cite{Zhang2025}, or data-driven approaches \cite{Liang2025}. This reconstruction framework reduces the pilot overhead associated with estimating all FA ports by inferring the spatial channel characteristics from a limited number of sampled FA positions. Note that this is important, as the pilot overhead scales the achievable rate as a prelog factor. Based on the reconstructed channels, the instantaneous SINR at all available ports is then obtained and used to select the port that provides the best received signal quality.

Noteworthy, most of the related literature \cite{New2025,Xu2025,Zhang2025,Liang2025,Kang2026} about channel estimation and reconstruction in FAS assumes a single-user scenario, even though the practical benefits of FAS are relatively limited in this situation compared to interference-limited multi-user setups. In this latter scenario \cite{Wong2023}, port selection involves searching for interference nulls rather than channel maxima. Hence, the objective shifts toward exploiting spatial positions that suppress multi-user interference and improve the SINR performance. In this case, estimating and reconstructing the channels of both the desired and the $U-1$ interfering users is needed to determine the SINR across all FA ports to perform port selection.

In spite of the similarity between slow-FAMA and conventional multi-user MIMO channel estimation strategies, a key difference should be emphasized. In conventional MIMO systems, accurate CSI is essential because significant performance gains can be obtained through precoding (and/or combining) and spatial multiplexing techniques. In contrast, slow-FAMA systems are inherently selection-based  because of their dependence on antenna port selection, and generally provide moderate gains through identifying the favorable antenna positions. This is exemplified in Fig. \ref{figure_2}, where the percentage rate loss due to channel estimation error is represented as a function of the normalized mean squared error (NMSE) for a reference zero-forcing (ZF) combiner (equalizer) compared to a FAMA-based alternative with the same overall aperture and same number of users. We see that to attain a rate loss of less than 20\%, ZF requires very accurate channel estimation to fully exploit the benefits of interference nulling; conversely, the same target is achieved by a FAMA scheme with a much lesser channel estimation quality. This suggests that, while highly accurate channel estimation in MIMO systems typically yields a large payoff, inheriting the same legacy procedures for slow-FAMA should, at the very least, be questioned.

%---------------------------------------------
% Section III - Myths and Misconceptions
%---------------------------------------------
\section{Myths \& Misconceptions in FAS Channel Estimation}
The previous section provided an overview of channel estimation and reconstruction in FAS. In this section, we delve into some myths that have been prevalent in the literature related to FAS in general, and to slow-FAMA in particular. These address the channel estimation and reconstruction problem from the angle of legacy MIMO systems, but a fundamentally different look is required to better adequate to the nature of FAS.

%--------------- Minimizing MSE is not optimal ---------------------------
\subsection{Estimation and reconstruction errors should be minimized}

\label{subsec:NMSE}

\begin{figure}[t]
\centering
{\includegraphics[width=8.5cm,height=6.2cm]{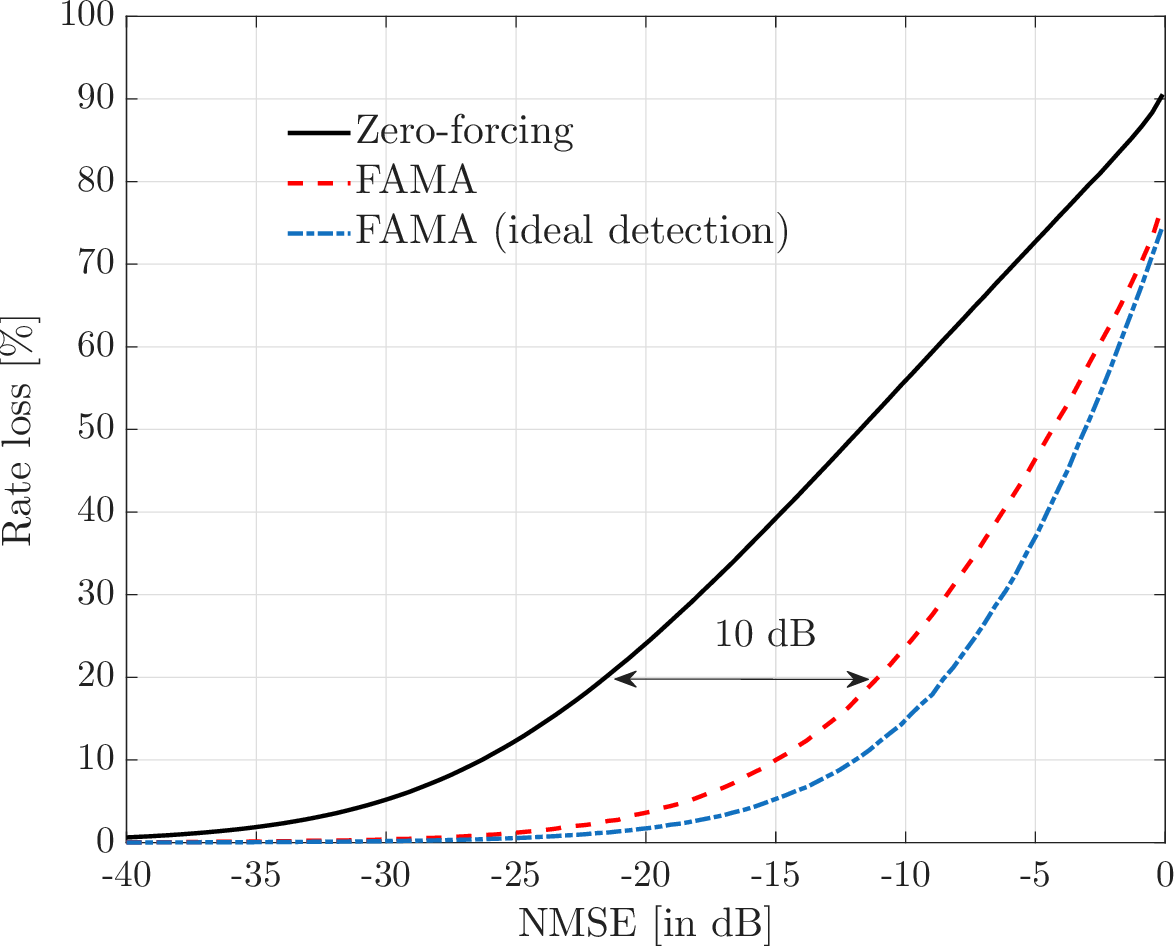}}
\caption{Rate loss versus NMSE for a zero-forcing combiner and a FAMA system under rich scattering (Rayleigh with Jakes' correlation). Minimum MSE channel estimation is assumed. $U=4$ users and a linear aperture of $W=5$ wavelengths are assumed in all instances.}
\label{figure_2}
\end{figure}

The FAS literature focuses on NMSE as the central metric to measure the accuracy of channel estimation and reconstruction. This is, mainly, due to the high impact of legacy MIMO systems in wireless research and the adoption of the same channel estimation procedures for FAS \cite{New2025,Xu2025,Zhang2025,Liang2025,Kang2026}. However, from Fig. \ref{figure_2} and the associated discussion, we observe that minimizing NMSE may be deceiving, at best.  

MIMO signal processing fully exploits amplitude and phase channel knowledge to construct beamformers which optimize the multiplexing capabilities and the radiation pattern. Hence, any error in the estimation leads to suboptimal beamformers and hence to a non-negligible performance loss. In turn, in FAMA systems the interest does not lie in a perfect channel reconstruction, but rather in accurate identification of the antenna positions rendering a maximum SINR. This leads to two important yet subtle observations (see Fig. \ref{figure_1}): \textit{i)} channel reconstruction may be very accurate and yield a rather low NMSE, but fail when reconstructing interference nulls -- which are precisely the ports of interest in slow-FAMA \cite{Wong2023}, and \textit{ii)} ports far away from the ``optimal" one (i.e., the one selected under perfect CSI) might yield very similar SINRs and, hence, very similar performance. Specifically, we can see in Fig. \ref{figure_1} that despite having a noticeable reconstruction error in the interfering channels associated with user 2 precisely at the interference nulls, and selecting a different port (blue square marker $\rm SP_{rec}$) than the optimal one under perfect CSI (teal square marker $\rm SP_{true}$), the operating SINR is barely affected.

Therefore, alternative metrics are needed that better capture the selection-based nature of FAMA systems. In particular, SINR loss and average rate loss (measured with respect to the optimal case in which perfect CSI is available) are more faithful metrics. Interestingly, rate loss naturally accounts for both the port selection error due to channel uncertainty and the associated pilot overhead, which scales rate as a prelog factor and may dominate in some regimes.

\begin{figure}[t]
\centering
{\includegraphics[width=8.5cm,height=6.2cm]{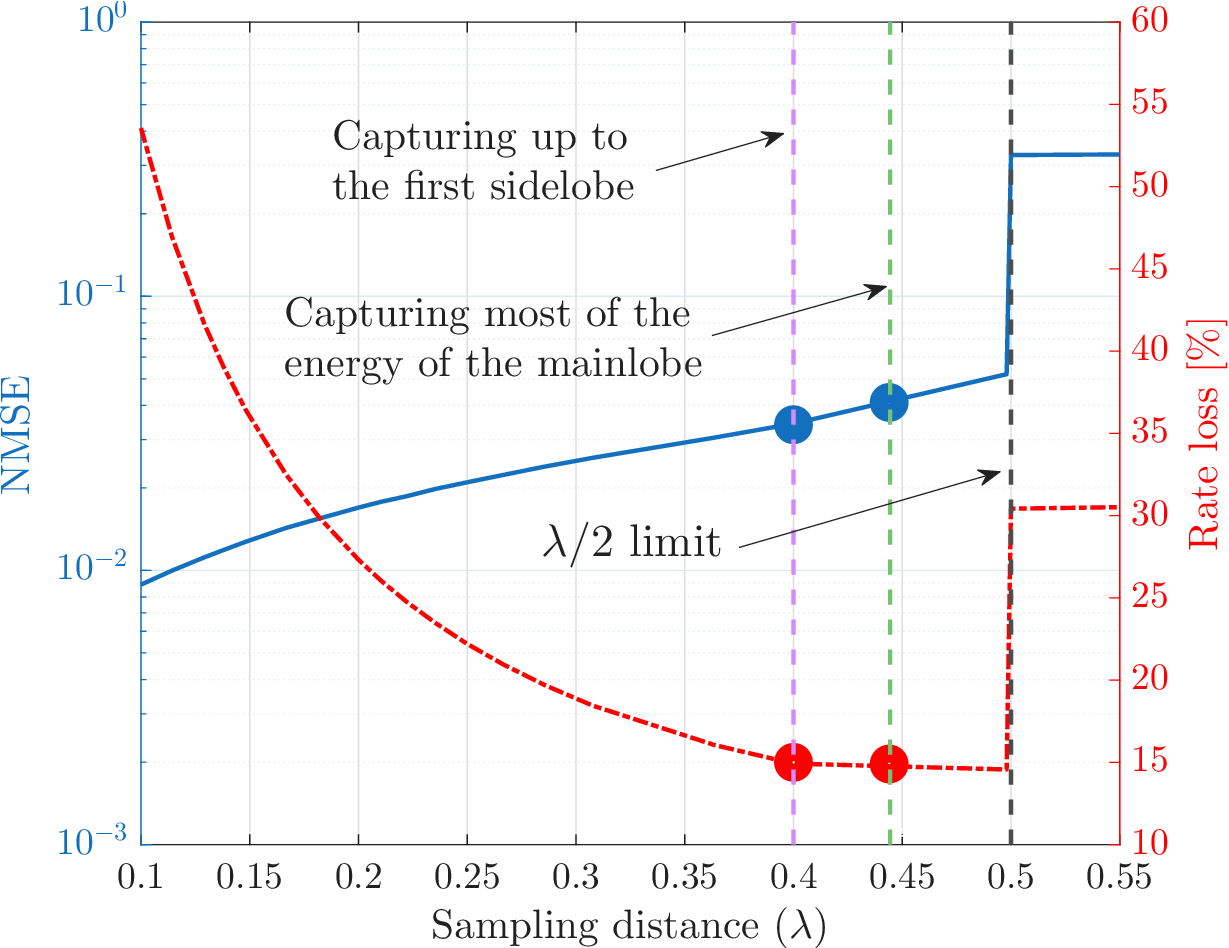}}
\caption{NMSE and rate loss versus sampling distance for a linear FA of size 8 wavelengths and 4 simultaneous users under rich scattering (Rayleigh with Jakes' correlation). The FA density is 20 ports per wavelength, and the average signal-to-noise ratio (SNR) is 10 dB.}
\label{figure_3}
\end{figure}

To illustrate this, Fig. \ref{figure_3} plots the NMSE and the average rate loss against the physical sampling distance. Note that, the smaller this distance, the more physical positions are sampled (i.e., a denser FA is considered), leading to a more accurate channel reconstruction at the cost of larger pilot overhead. The considered setup employs Gaussian processes (GP)-based reconstruction, where the kernel is estimated offline with 1000 training samples. Coherence blocks of 1200 digital symbols are assumed, with $2U$ pilots for channel estimation. As can be observed, Fig. \ref{figure_3} reveals a key insight: while NMSE naturally decreases with smaller sampling distances (implying a larger $N_{\rm S}$), such an increase in channel estimation accuracy does not translate into a lower rate loss. In fact, once the channel reconstruction quality is \textit{sufficient}, the increased number of pilots due to the extra sampling positions penalizes the achievable rate. In other words, the diminishing returns of rate gain due to finer channel estimation do not compensate for the extra pilot overhead. Importantly, observe that a sampling distance \textit{slightly below} half-wavelength (the limit dictated by Nyquist sampling) suffices to fully exploit FAS potential. Capturing additional sidelobes of spectral leakage inherent to the finite aperture of FAS as suggested in \cite{New2025} may not be needed when the focus is not on improving the NMSE \textit{per se}.

%--------------- Estimating channels vs SINR ---------------------------
\subsection{Estimating channels or estimating SINR?}

Building on the previous discussions, it is apparent that the ultimate metric of interest in slow-FAMA systems is the SINR at all available ports. This metric can be obtained by estimating and reconstructing the desired and the $U-1$ interfering channels in the pilot stage (see Fig. \ref{figure_1}), to compute the reconstructed SINR at the $N$ ports. An accurate estimation and reconstruction of the desired channel is advisable for signal detection, and in practice it can almost be taken for granted as the SINR at the selected port is associated to a good quality for the signal of interest. However, an accurate estimation and reconstruction of the individual channels of all interfering users at all ports is costly in terms of system and receiver complexity, and maybe not needed at all.

This motivates reconstructing the SINR directly across the available antenna ports from SINR samples obtained at the $N_{\rm S}$ sampling positions, instead of estimating and reconstructing all $U$ channels at all $N$ ports and then computing the SINR at all ports. In this reconstruction method, the SINR can be treated as the target random process, and an interpolation method---such as filtering \cite{New2025} or GP \cite{Zhang2025}---can be directly applied to a limited set of SINR sampled observations. Alternatively, one can estimate the desired channel as conventionally, and the aggregate interference-plus-noise power separately \cite{Dinis2026}. Particularly, the estimation of the aggregate interference-plus-noise is first obtained by subtracting the contribution of the estimated desired signal from the received observations, followed by separate reconstruction of both the desired signal and the aggregate interference-plus-noise power field across all FA ports. The reconstructed SINR is then computed by dividing the reconstructed desired signal power over the reconstructed aggregate interference-plus-noise power. With this method, which we refer to as aggregate interference-aware (AgIA) reconstruction, SINR can be reconstructed with the same complexity regardless of the number of interfering users. This is particularly relevant, since state-of-the-art evolutions of FAMA envision $\times10$ multiplexing capabilities, which implies a much larger number of interfering signals.

\begin{figure}[t]
\centering
{\includegraphics[width=8.5cm,height=6.2cm]{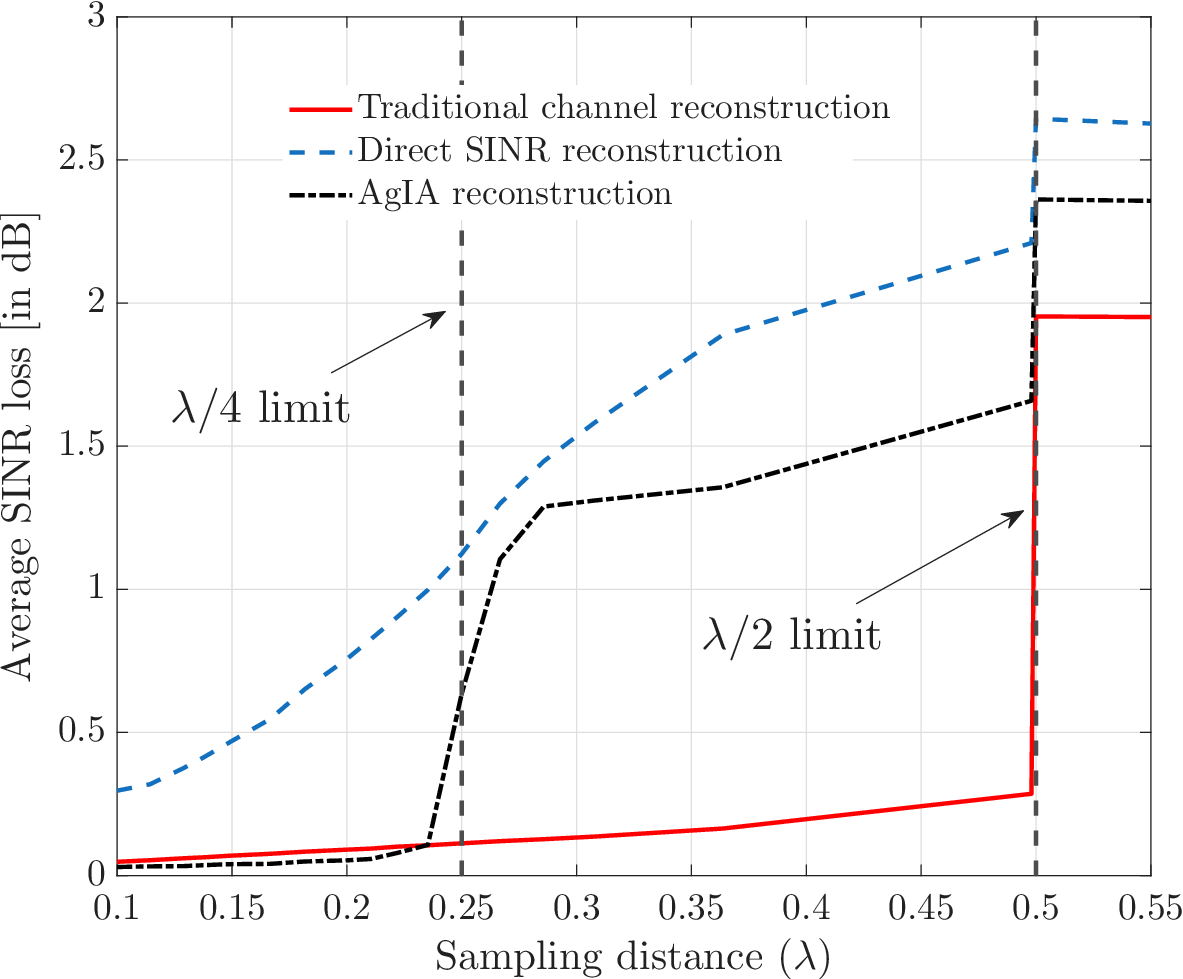}}
\caption{Average SINR loss versus sampling distance, assuming four users, a FA of size $W=8$ wavelengths, and a FA density of 20 ports/wavelength.}
\label{figure_4}
\end{figure}

A comparison of the three approaches---traditional estimation of the $U$ channels, direct SINR reconstruction, and AgIA reconstruction---is presented in Fig. \ref{figure_4}, where the SINR loss is depicted in terms of the sampling distance. Focusing first on the traditional channel estimation procedure (each channel is estimated separately as in MIMO systems), we observe that the performance drop is almost negligible for sampling distances \textit{slightly below} half-wavelength ($\lambda/2$) onward, which is coherent with the statistics of the windowed spatial channel process \cite{New2025}. Remarkably, for the AgIA reconstruction, this turning point shifts towards $\lambda/4$; it can be formally proved\footnote{The proof is not included due to the specific format of magazine submissions, but can be derived from a wavenumber domain formulation of power magnitudes similar to the one used in \cite{New2025} for signal amplitudes.} that when reconstructing a power-domain measure (e.g., the aggregate interference), the theoretical sampling limit approaches a quarter wavelength. In other words, power fields in the AgIA reconstruction method require sampling slightly below $\lambda/4$, instead of $\lambda/2$ as in conventional channels. This important conclusion allows us to move beyond legacy MIMO estimation procedures, considerably reducing the receiver complexity at the cost of a non-drastic denser sampling process. 
%\pre{Complete once the figure includes also the different spatial correlation curves, and when the inconsistency with the port selection error in Fig. 6 is clarified (direct SINR reconstruction has similar port selection error but awful performance)}

%--------------- Oversampling is essential ---------------------------
\subsection{Oversampling is essential}

The concept of oversampling is widely interpreted as using sampling distances smaller than the conventional Nyquist limit, which in the spatial domain relates to half-wavelength ($\lambda/2$) sampling. In the FAS literature, assuming that $N$ positions or \textit{ports} are available for spatial sampling along the FA dimension of $W$ wavelengths, finding the minimum number of samples $N_{\rm S}<N$ required to reconstruct the spatial random process with enough precision nicely connects with sampling theory: the number of dominant eigenvalues (e.g. eigenvalues containing over 99\% energy) of the channel spatial correlation matrix is roughly the number of half-wavelengths contained in the FA aperture, i.e. $N_{\rm S}\approx\lceil 2 W\rceil $ \cite{Ramirez2025}. This is consistent with the results in \cite{New2025} obtained for a single-user case, where the need for spatial oversampling (i.e., sampling below what's predicted by Nyquist law $\lambda/2$) for a satisfactory signal reconstruction is stated. Specifically, the number of samples across a given dimension needed to capture the spectrum energy up to the $d$-th sidelobe is approximately $N_{\rm S}\approx\lfloor 2 (W+d+1)\rfloor$, or, similarly, the minimum (sub-optimal) sampling distance is given by $D^*\approx \frac{W\lambda}{2(W+d+1)}$, trading off spatial resolution and reconstruction accuracy through $d$. However, while this oversampling may be desirable from an NMSE perspective, the results in Fig. \ref{figure_3} suggest that the need for oversampling can be relaxed, as a couple of extra spatial samples beyond the conventional Nyquist sampling half-wavelength distance seem sufficient. This leads to a key question regarding how the very notion of oversampling should be defined in the context of slow-FAMA systems.

It is worth noting that the required sampling distance inherently depends on the physical magnitude being reconstructed, as noticed from Fig. \ref{figure_4}. While reconstructing the complex channel field itself in the traditional channel reconstruction method is governed by the classical $\lambda/2$ sampling criterion, reconstructing the aggregate interference-plus-noise power requires a $\lambda/4$ sampling condition. In this situation, though, sampling slightly under $\lambda/4$ should not be formally regarded as oversampling, as just \textit{a few extra samples} beyond the theoretical limit are required.

Another important observation is that the above sampling conditions are all derived from assuming isotropic propagation conditions (Jakes' spatial correlation model). However, this is actually the \textit{worst-case} scenario for overhead. In this theoretical limit case, the spatial correlation is at its minimum, requiring the maximum number of samples within the aperture to capture the signal's degrees of freedom.

\begin{figure}[t]
\centering
{\includegraphics[width=8.5cm,height=6.2cm]{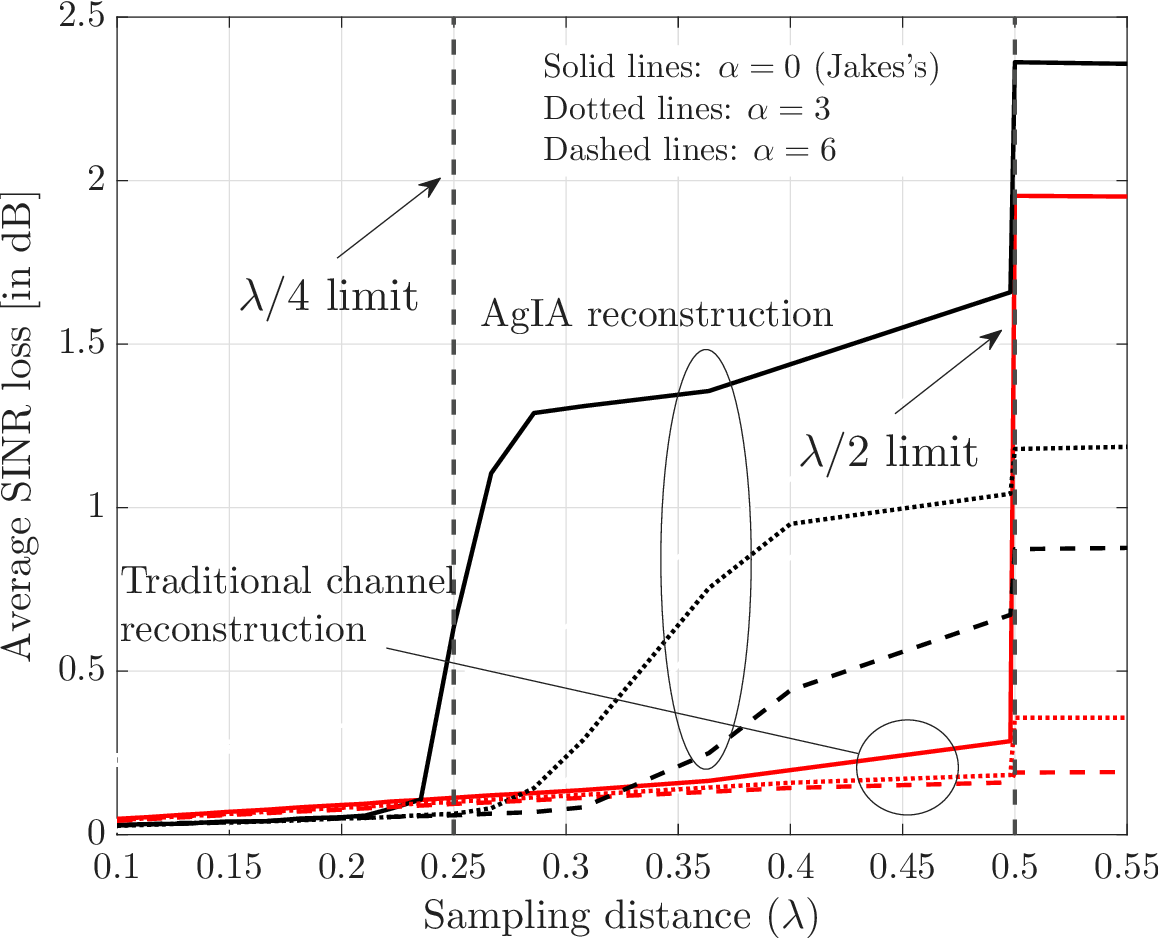}}
\caption{Average SINR loss versus sampling distance with different correlation functions, assuming 4 users, a FA of size $W=8$ wavelengths, and a FA density of 20 ports/wavelength.}
\label{figure_5}
\end{figure}

In real-world non-isotropic scenarios (e.g., clustered scattering), the spatial rank of the channel is significantly lower, which implies less dominant eigenvalues and hence a smaller number of sampling points $N_{\rm S}$. This inherent sparsity can be exploited to further reduce sampling complexity and simplify the reconstruction process, making FAS even more efficient in practical deployments than in theoretical models. To illustrate this, Fig. \ref{figure_5} plots the SINR loss in terms of the sampling distance for different spatial correlation structures. Specifically, an angular distribution of the type $\sin(\cdot)^\alpha$ is used, where the larger $\alpha$ the more concentrated the angles of arrival and thus the less dominant eigenvalues of the channel spatial correlation. As predicted, the sampling distance at which the performance loss remains negligible increases as the spatial diversity is reduced ($\alpha$ increases). This is specially noticeable for the AgIA reconstruction method, where the turning point for negligible performance loss moves closer to $\lambda/2$ as $\alpha$ grows. Therefore, isotropic scattering (Jakes' spatial correlation) is useful as worst-case design and as baseline to compare different channel estimation approaches. In practice, though, more favorable conditions (from overhead viewpoint) can be expected, and be used dynamically to improve system performance.

\subsection{Port selection accuracy is key}
\begin{figure}[t]
\centering
{\includegraphics[width=8.5cm,height=6.2cm]{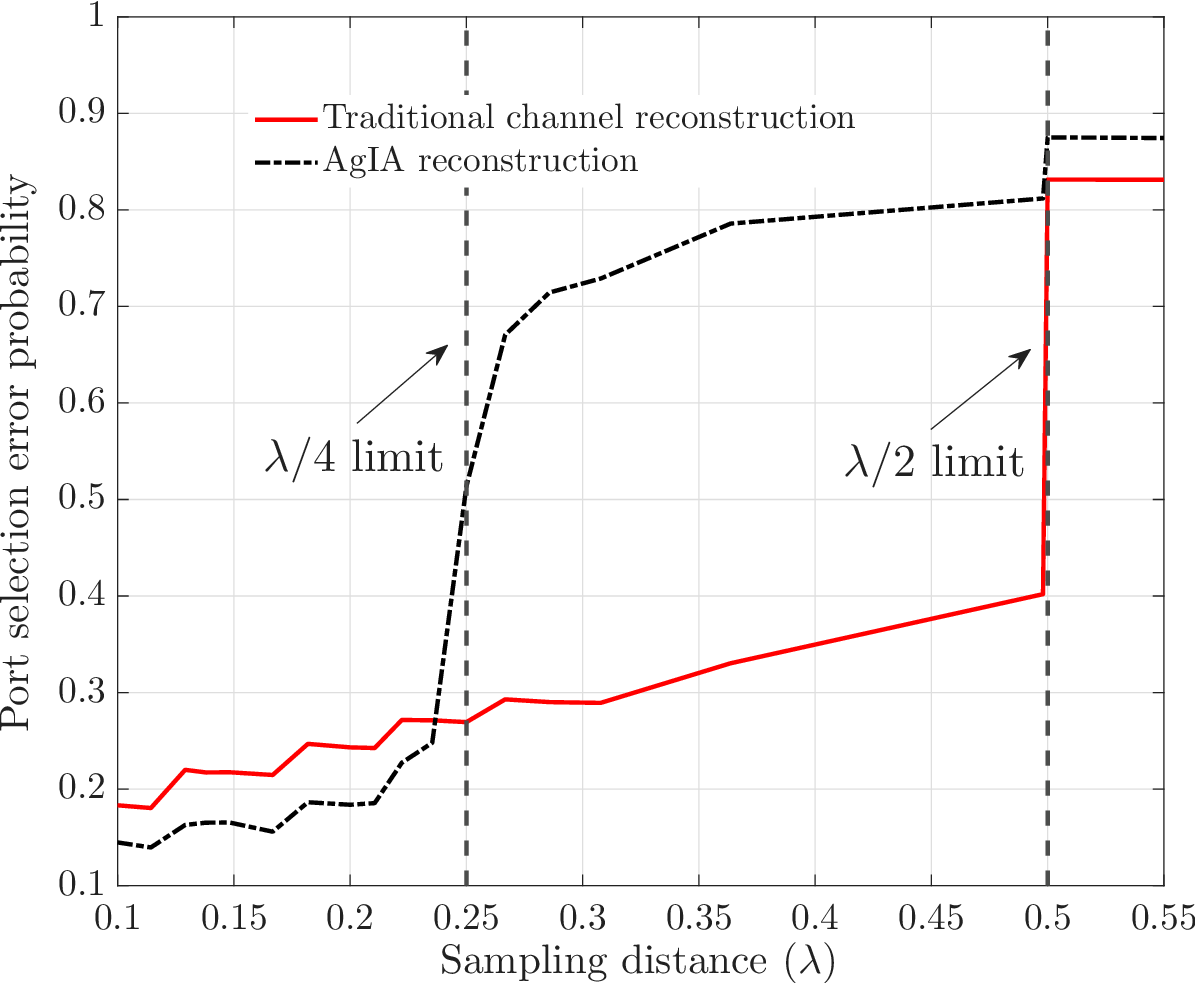}}
\caption{Port selection error probability versus sampling distance, assuming four users, a FA of size W = 8 wavelengths, and a FA density of 20 ports/wavelength.} 
\label{figure_6}
\end{figure}

The open-loop nature of FAMA shifts the need to acquire CSI to the receiver ends, which is required for port selection and signal detection. While there is abundant literature dedicated to the port selection problem in FAS \cite{Chai2022}, a formal definition of what constitutes wrongful port selection has not been established in the literature to the best of our knowledge. In Fig. \ref{figure_6}, we use a hard definition\footnote{We define an error event as selecting a different port of the optimal one under perfect CSI knowledge.} of port selection error to understand its interplay with the sampling distance and the reconstruction procedure. The joint interpretation of Figs. \ref{figure_4} and \ref{figure_6} shows some interesting insights: (\emph{i}) FAMA operate well even under 20-30\% port selection error---this shows their robustness against estimation inaccuracies; (\emph{ii}) below the quarter-wavelength limit, interference-aware reconstruction yields a slightly improved accuracy than traditional channel reconstruction in terms of port selection success; (\emph{iii}) port selection error alone is not a proxy for performance degradation in FAS.

In fact, from Fig. \ref{figure_1} and the associated discussion in Section \ref{subsec:NMSE}, we can even wonder whether port selection error definitions should build on the distance to the optimal port or on performance loss due to port selection mismatch, as far away ports might yield very similar performance. Moreover, for dense FAs, a relatively large number of close-by ports yield similar performance, as the physical distance between them is sufficiently low that the channel (more importantly, SINR) experiences very high correlation, virtually representing the same effective port.

\section{Some Critical Questions}
Having exposed the pitfalls of applying legacy MIMO paradigms to inherently selection-based systems as FAS, a performance-driven shift is required. This section outlines four critical questions that must be answered to define a concrete roadmap for practical FAMA deployment.

\subsection{Question 1: What is the Sampling Law for Port Selection?}
Nyquist theory for spatial sampling establishes a limit to the minimum number of samples required to reconstruct the signal over the FA aperture. Depending on whether complex signals or power-domain metrics are targeted, and the spatial correlation profile, Nyquist limit can be different. We have seen that despite oversampling may be beneficial to reduce estimation error \cite{New2025}, such condition can be relaxed when focusing on achievable performance. However, a meaningful definition of sampling law and port selection error that reconciles error and performance metrics is yet to be investigated: how many samples $N_{\rm S}$ are required for port selection for an $\epsilon$ degradation in achievable performance?% remains an open question. The use

%Nyquist-Shannon tells us how many samples are required to reconstruct a bandlimited field. However, FAMA requires a different theoretical foundation, i.e., sort of a \textit{sampling theorem for port selection}. Specifically, how many samples $N_{\rm S}$ are required for port selection with some $\epsilon$ degradation in achievable performance? This shifts the problem from signal reconstruction to stochastic spatial search. \pre{This question is partially answered alrady. However, an interesting thing is exploring how changing the spatial correlation impacts the SINR minimum sampling distance. }

\subsection{Question 2: What is the proper method for SINR reconstruction}

Classical methods such as low-pass filtering \cite{New2025} have shown good performance for channel reconstruction in point-to-point links. Bayesian learning approaches offer a mathematically robust framework that can be used to improve estimation \cite{Xu2025,Zhang2025}. Shifting the estimation and reconstruction problem to the power domain may require twice as many samples but makes complexity independent of the number of interfering signals. This opens the door to implementing refined kernel estimation mechanisms, provided that they are able to operate within the channel coherence interval.

%While filtering methods are ill-suited for the non-linear topography of SINR, GPR offers a mathematically robust framework, provided that we can identify the correct kernels. Conversely, while learning-based methods are promising, their reliance on frequent retraining in dynamic environments remains a significant issue. Is a model-aware GPR with a ground-truth kernel the ultimate solution, or instead will learning techniques prevail? 

\subsection{Question 3: How do we balance multi-port sensing vs. selection gain?}
As hardware moves toward multi-port embodiments of FAS, we face a new fundamental trade-off: is it more efficient to utilize multiple ports to sense the environment at a lower resolution simultaneously, or to use a single port to sense at a higher resolution sequentially? Solving this will require integrating multi-port spatial correlation and coupling into  channel estimation and reconstruction protocols \cite{Coma2026,Dinis2026}.

\subsection{Question 4: How do the previous results translate to electronically reconfigurable FAS?} 
The advent of the tri-hybrid MIMO architecture \cite{Heath26} extends the legacy hybrid MIMO implementations to leverage the beampattern flexibility of electronically reconfigurable antennas \cite{Liu2026}. The ability of metasurface-based embodiments \cite{Liu2025, Zhang2026} to emulate the behavior of conceptual FAs paves the way to an architectural convergence between multiport FA solutions, tri-hybrid MIMO and general metasurface antennas, requiring the design of new channel estimation and reconstruction protocols that are able to fully exploit the benefits and flexibility of these implementations. Although the core principle remains---estimating spatially (beamspace) correlated processes from a few samples---the different FA designs might impose new challenges and fundamental limits.
%\textcolor{blue}{I think it is important to pose this question, generalize the channel estimation problem as selecting the optimal port (either virtual or physical) from a large set of correlated samples drawn from a likely unknown random process.}

\section{Conclusions}

In this work, we provided a critical look at the channel estimation and  reconstruction problem in the context of multi-user communications under the slow-FAMA paradigm. As a key novelty compared to legacy MIMO systems, we argued that in FAMA one does not need to sacrifice resources for the sake of obtaining refined channel estimates, as spatial oversampling quickly yields diminishing returns in terms of performance improvement. To fully exploit the potential of FAMA, one must abandon pursuing the holy grail of minimizing global error metrics, and look elsewhere: spatial oversampling demands can be significantly relaxed when considering performance-driven selection metrics, and aggregate interference-aware reconstruction offers a highly scalable, low-complexity path forward.

Looking ahead, we identify four critical question that deserve special attention to further develop FAMA into a solid candidate for open-loop multiple access in next-generation technologies: performance-oriented sampling laws, power-domain reconstruction methods, multi-port sensing trade-offs, and the transition toward electronically reconfigurable antenna surfaces and tri-hybrid architectures. These are key to unlocking the true multiplexing and interference-nulling capabilities of FAS as an alternative to closed-loop MIMO solutions.

% near-perfect accuracy with small reconstruction errors may have a negligible influence on the port selection decision, 

% These metrics reflect directly the effect of channel estimation and reconstruction on the rate performance, rather than on the reconstruction accuracy.

% =====================
\section*{Acknowledgments}
This work is supported by grants PID2023-149975OB-I00 (COSTUME), PREP2023-001949 and RYC2020-030536-I funded by MICIU/AEI/10.13039/501100011033 and FEDER/UE, and by the European Union under the Marie Sklodowska-Curie grant agreement No. 101109529. 
\bibliographystyle{IEEEtran}
\bibliography{refs}

% =====================
%\section*{Biographies}

\begin{IEEEbiographynophoto}{Taissir Y. Elganimi} [SM] (etaissir@ugr.es) is a predoctoral researcher at University of Granada. His research interests include reconfigurable intelligent surfaces for wireless communications, and fluid antenna systems.
\end{IEEEbiographynophoto}

\begin{IEEEbiographynophoto}{P. Ram\'irez-Espinosa} (pabloramirez@uma.es) is an MSCA postdoctoral fellow at University of M\'alaga. His research interests include dynamic metasurfaces antennas, fluid antennas, and signal processing for communications.
\end{IEEEbiographynophoto}

\begin{IEEEbiographynophoto}{D. Morales-Jim\'enez}
[SM] (dmorales@ugr.es) is an RyC Research Professor at University of Granada. His research interests include statistical signal processing, random matrix theory, and high-dimensional statistics.
\end{IEEEbiographynophoto}

\begin{IEEEbiographynophoto}{F. Javier López-Martínez}
[SM] (fjlm@ugr.es) is a Professor at University of Granada. His research interests include dynamic reconfigurable arrays for wireless communications, signal processing, and physical layer security.
\end{IEEEbiographynophoto}

\end{document}